\documentclass{ws-procs975x65}
\usepackage{graphicx}

\def\beq{\begin{equation}}
\def\eeq{\end{equation}}

\begin{document}

\title{Magnetized $1.5$-dimensional advective accretion flows around black holes}
\author{Tushar Mondal$^*$ and Banibrata Mukhopadhyay$^\dagger$}

\address{Department of Physics, Indian Institute of Science,
	Bengaluru 560012, India\\
$^*$E-mail: mtushar@iisc.ac.in\\
$^\dagger$E-mail: bm@iisc.ac.in}

\begin{abstract}
We address the role of large scale magnetic stress in the angular momentum transport, as well as the formation of different kinds of magnetic barrier in geometrically thick, optically thin, vertically averaged 1.5-dimensional advective accretion flows around black holes. The externally generated magnetic fields are captured by the accretion process from the environment, say, companion stars or interstellar medium. This field becomes dynamically dominant near the event horizon of a black hole due to continuous advection of the magnetic flux. In such magnetically dominated accretion flows, the accreting matter either decelerates or faces magnetic barrier in vicinity of the black hole depending on the magnetic field geometry. We find that the accumulated strong poloidal fields along with certain toroidal field geometry help in the formation of magnetic barrier which may knock the matter to infinity. When matter is trying to go back to infinity after getting knocked out by the barrier, in some cases it is prevented being escaped due to cumulative action of strong magnetic tension and gravity, and hence another magnetic barrier. We suggest, this kind of flow may be responsible for the formation of episodic jets in which magnetic field can lock the matter in between these two barriers. We also find that for the toroidally dominated disc, the accreting matter rotates very fast and decelerates towards the central black hole.
\end{abstract}


\bodymatter


\section{Introduction}

 In 1972, Bekenstein proposed an idealized engine, namely Geroch-Bekenstein engine \cite{B72}, that makes use of the extreme gravitational potential of a black hole (BH) to convert mass to energy with almost perfect efficiency. However, the practical realization of this engine is very difficult in astrophysical systems. Generally astrophysical BHs do convert mass to energy via accretion process with modest efficiencies. Later the idea came to address the importance of large scale dipole magnetic field in an accretion flow \cite{BK74}. Following this idea, it was suggested and also verified numerically \cite{TNM11} that this efficient conversion is practically possible in the presence of large scale poloidal magnetic field. Such an efficient accretion phenomenon is named as Magnetically Arrested Disc (MAD) \cite{NIA03}.
 
 The origin of strong magnetic field near the event horizon of a BH is as follows. For the case of advective accretion flow, the magnetic fields are captured from accreting medium or companion star. These fields are dragged inward with continuous accretion process and become dynamically dominant through flux freezing in vicinity of a BH. Theoretical models suggest the importance of large-scale magnetic field in accretion and in the formation of strong outflows/jets, as well as high-energy radiation \cite{MKU01, MM19a, MM19b}. Observations of all well-known jetted sources also indicate the presence of dynamically important magnetic field at the jet-footprint \cite{Z14}. In this proceeding, we explore the underlying role of such large-scale strong magnetic fields on the disc flow behaviours for optically thin, geometrically thick, advective accretion flows around BHs. Unlike MAD, here the advection of both poloidal and toroidal magnetic fields are considered. We address the possible origin of different kinds of magnetic barriers depending on field geometry.
 
\section{Model equations} \label{sec:model}
We address magnetized, optically thin, advective, axisymmetric, vertically averaged, steady-state accretion flow around BHs in the pseudo-Newtonian framework with Mukhopadhyay potential \cite{B02}. The flow parameters, namely, radial velocity $(v)$, specific angular momentum $(\lambda)$, fluid pressure $(p)$, mass density $(\rho)$, the radial $(B_{r})$ and toroidal $(B_{\phi})$ components of magnetic fields, are solved simultaneously as functions of radial coordinate $r$. Throughout in our computation, we express the radial coordinate in units of $r_{g}=GM_{BH}/c^{2}$, where $G$ is Newton's gravitational constant, $M_{BH}$ is the mass of the black hole, and $c$ is speed of light. We also express $\lambda$ in units of $GM_{BH}/c$ and velocity in units of $c$, and the other variables accordingly to make all the variables dimensionless. Hence, the continuity equation, the components for momentum balance equation, and the energy equation can read as, respectively,
\begin{eqnarray}
&\frac{d}{dr}\left(r\rho h v\right)=0 ,\label{eq:continuity}& \\ &v\frac{dv}{dr}-\frac{\lambda^{2}}{r^{3}}+\frac{1}{\rho h}\frac{d(hp)}{dr}+F=-\frac{\sqrt{h} B_{\phi}}{4 \pi \rho h r}\frac{d}{dr}\left(r\sqrt{h}B_{\phi}\right),\label{eq:radmomentum}& \\ &v\frac{d\lambda}{dr}=\frac{1}{r\rho h}\frac{d}{dr}\left(r^{2}W_{r\phi}h\right)+\frac{\sqrt{h}B_{r}}{4 \pi \rho h}\frac{d}{dr}\left(r\sqrt{h} B_{\phi} \right),\label{eq:angmomentum}& \\ &\frac{hv}{\Gamma_{3}-1}\left(\frac{dp}{dr}-\frac{\Gamma_{1} p}{\rho}\frac{d\rho}{dr}\right)=Q^{+}-Q^{-}=f_{m}Q^{+}=f_{m}(Q^{+}_{vis}+Q^{+}_{mag}), \label{eq:energy}&
\end{eqnarray}
where $F$ is the magnitude of the gravitational force corresponding to the pseudo-Newtonian potential. $W_{r\phi}$ is the viscous shearing stress, which can be expressed using Shakura-Sunayev \cite{SS73} $\alpha$-viscosity prescription with appropriate modifications \cite{C96} due to advection as given by  $W_{r\phi}=\alpha (p+\rho v^{2})$. Following the vertical momentum balance equation, the disc half-thickness can be written as
\begin{eqnarray}
h= r^{1/2} F^{-1/2} \sqrt{\left(p+\frac{B^{2}}{8\pi}\right)\big/\rho}.
\end{eqnarray}
The energy equation is written by taking care of the proper balance of heating, cooling and advection. $Q^{+}$ is the energy generated per unit area due to magnetic $(Q^{+}_{mag})$ and viscous dissipation $(Q^{+}_{vis})$, whereas $Q^{-}$ infers the energy radiated out per unit area through different cooling mechanisms \cite{MM19b}. However for the present purpose, we do not incorporate cooling processes explicitly. The factor $f_{m}$ measures the degree of cooling and it varies from 0 to 1 for two extreme cases of efficient cooling and no cooling respectively. The details of dissipation terms are followed from the Ref.~\refcite{MM18}. Note that in the energy equation, we do not include the heat generated and absorbed due to nuclear reactions \cite{DM19}.

The presence of magnetic field provides two other fundamental equations, namely, the equation for no magnetic monopole and the induction equation. These are respectively
\begin{equation}
\nabla.\mathbf{B}=0, \ \text{and} \ \nabla \times \left(\mathbf{v}\times\mathbf{B} \right)+\nu_{m}\nabla^{2}\mathbf{B}=0, \label{eq:induction}
\end{equation}
where $\mathbf{v}$ and $\mathbf{B}$ are the velocity and magnetic field vectors respectively and $\nu_{m}$ is the magnetic diffusivity. For this accretion disc solution, we consider the induction equation in the limit of very large magnetic Reynolds number $(\propto 1/\nu_{m})$. 

\section{Solution procedure}
The set of six coupled differential equations $(1)-(6)$ for six flow variables $v$, $\lambda$, $p$, $\rho$, $B_{r}$, and $B_{\phi}$ are solved simultaneously to obtain the solutions as functions of the independent variable $r$. The appropriate boundary conditions are as follows. Very far away from the BH, matter is sub-sonic and the transition radius between the Keplerian to sub-Keplerian flow is the outer boundary of our solutions. Very near the BH, matter is super-sonic and the event horizon where matter velocity reaches speed of light is the inner boundary. In between these two boundaries, matter becomes tran-sonic where matter velocity is equal to (or similar to) medium sound speed and this location is called as sonic/critical point. We use this point as one of the boundary. Different types of sonic/critical points, say, saddle, nodal, and spiral, are described in the Ref.~\refcite{MM18}.

\section{Results}

The angular momentum transport in accretion physics had been a long standing issue until Balbuas \& Hawley \cite{BH98} described the importance of magnetorotational instability (MRI), particularly in an ionized medium in the presence of weak magnetic fields. Later Mukhopadhyay \& Chatterjee \cite{MC15} showed that the efficient transport of angular momentum is also possible by large-scale magnetic field in geometrically thick, advective accretion flow, even in the complete absence of $\alpha$-viscosity. Here, we address the importance of large-scale strong magnetic field in accretion geometry. When the magnetic field is strong enough, presumably that corresponds to the upper limit to the amount of magnetic flux which a disc around a BH can sustain, it disrupts the accretion flow. Then the force associated with the magnetic stress becomes comparable to the strong gravitational force of the BH. As a consequence, the magnetic field arrests the infalling matter in vicinity of the BH. Here we plan to understand such magnetic activity depending on different kinds of field geometry. 

\begin{figure}[h]
	\begin{center}
		\includegraphics[width=\columnwidth]{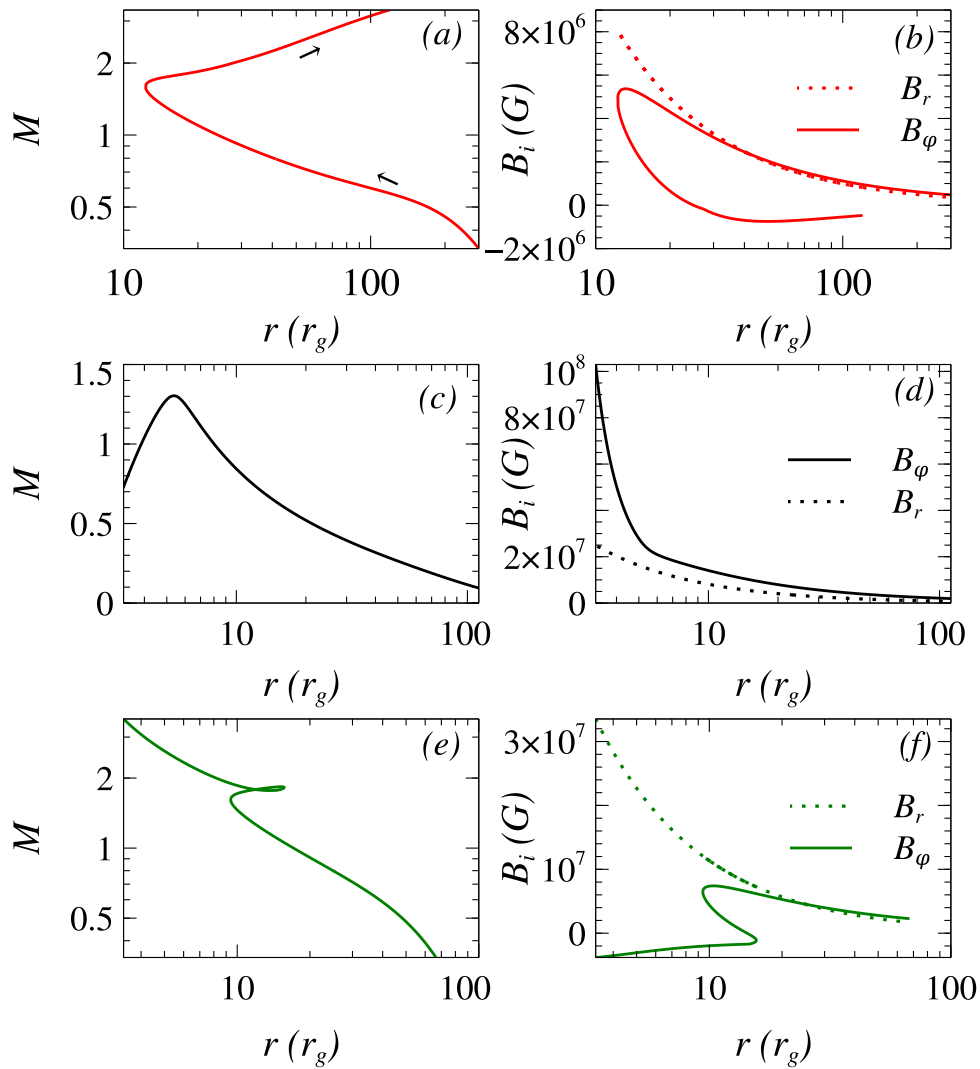}
	\end{center}
	\caption{Variation of Mach number $(M)$ and the corresponding magnetic field components $(B_{i})$ for different relative field strengths in the disc. The model parameters are $M_{\text{BH}}=10M_{\odot}$, $\dot{M}=0.01\dot{M}_{\text{Edd}}$ and $f_{m}=0.5$.}
	\label{fig:fig1}
\end{figure}

In Figure~\ref{fig:fig1} we show three different natures of accretion depending on different magnetic field geometries. The left column indicates the Mach number $(M)$, defined as the ratio of the radial velocity to the medium sound speed, whereas the right column represents the corresponding magnetic field components $(B_{i})$. 
In Figure~\ref{fig:fig1} $(a)$, the Mach number increases monotonically towards the central BH until it faces magnetic barrier at $r\simeq 12$. After facing such a barrier, matter goes back to infinity. To explain the origin of such barrier, we focus on the underlying role of magnetic force along with strong gravity, centrifugal force, and force due to gas pressure. When matter falls towards central BH, the forces supporting gravity may suppress by the forces acting outward. In such circumstances, following radial momentum balance equation, the condition for which barrier appears as
\begin{eqnarray}
	-\frac{1}{4\pi \rho h}\left(\sqrt{h}B_{\phi}\frac{d(\sqrt{h}B_{\phi})}{dr}\right)-\frac{1}{\rho h}\frac{d(hp)}{dr}+\frac{\lambda^{2}}{r^{3}}>F+\frac{1}{4\pi \rho h}\frac{(\sqrt{h}B_{\phi})^{2}}{r}, \label{eq:barrier}
\end{eqnarray}
where the first term appears from magnetic pressure and the last term from magnetic tension. Simply it suggests the essential condition to appear magnetic barrier is $\frac{d}{dr}(\sqrt{h}B_{\phi}) \ll -\frac{\sqrt{h}B_{\phi}}{r}$. On the other hand, the equation for no magnetic monopole leads to $\frac{d}{dr}(\sqrt{h}B_{r}) = -\frac{\sqrt{h}B_{r}}{r}.$ These combine conditions suggest that the matter faces magnetic barrier only when the field strength is strong enough to satisfy equation~(\ref{eq:barrier}) and the disc is poloidal magnetic field dominated. Figure~\ref{fig:fig1}$(b)$ infers such field conditions.

In Figure~\ref{fig:fig1}$(c)$, the Mach number initially increases monotonically and finally slows down to the central BH. In this case, the disc is toroidal magnetic field dominated as shown in Figure~\ref{fig:fig1}$(d)$. In this situation matter is rotating very fast compared to the inward dragging. In Figure~\ref{fig:fig1}$(e)$, the Mach number increases up to $r\simeq 9.5$, where the magnetic barrier appears due to dominant poloidal magnetic field. After knocking out by such a barrier, matter tries to go away from the BH, but faces another barrier at $r\simeq 15.6$. This is because of the cumulative action of the inward strong gravity force and the inward force due to magnetic tensions along the field lines. In between these two barriers matter faces spiral-type critical point and again falls back to the central BH. The corresponding field components are shown in Figure~\ref{fig:fig1}$(f)$. Note that the magnetic field strength within plunging region is few factor times $10^{7}$ G for $10 M_{\odot}$ BH for all the cases mentioned above.

\section{Discussions}

Accretion disc can carry small as well as large-scale strong magnetic fields. Small-scale field may generate, locally, through some physical scenario, Biermann batery mechanism, dynamo process etc. However, the origin of large-scale strong magnetic fields is still not well understood. It was suggested that the externally generated magnetic field either from companion star or from interstellar medium may capture and drag inward through continuous accretion process. This field becomes dynamically dominant in vicinity of a BH due to flux freezing. In such circumstances, strong magnetic fields arrest the infalling matter and disrupt the conventional accretion flow properties.
 
In this proceeding, we address three different types of accretion flow properties depending on field geometry in the presence of large scale strong magnetic fields. First, when the disc is poloidal field dominated, the infalling matter faces magnetic barrier and tries to go back to infinity. On its way away from the BH, matter may completely lose its angular momentum and at the same time the combined effects of inward strong gravity and the magnetic tension along field lines can prevent the matter being escaped. In this particular case, matter again falls back to the BH by facing second magnetic barrier. The important application of such a flow behaviour is the formation of episodic jets in which magnetic field can lock the infalling matter in between these two magnetic barriers. Second, for such a poloidal field dominated case, matter can completely go back to infinity after knocked out by the first magnetic barrier. Magnetic tension is not strong enough to prevent the matter being escaped in this case. This type of flows is the building block to produce unbound matter and hence strong continuous outflows/jets. Third, when the disc is toroidal field dominated, the infalling matter rotates very fast rather than being dragged inward in vicinity of the BH. In such case, matter falls slowly into the BH. For all these scenarios the maximum magnetic field strength near the stellar mass BH is few factor times $10^{7}$ G, which is well below the Eddington magnetic field limit \cite{MM19a} and hence perfectly viable.
The other important aspect of this quasi-spherical advective flow is the angular momentum transport. The outward transport of angular momentum occurs through large-scale magnetic stress. The specific angular momentum is quite below the local Keplerian value.

In the other proceedings of this volume, we discuss more general disc-outflow symbiotic model \cite{MMMG15} and its observational implication to ultra-luminous X-ray sources \cite{MMG15}.

\section*{Acknowledgments}

TM acknowledges P. V. Lakshminarayana travel grant ODAA/INT/18/11, Office of Development and Alumni Affairs, Indian Institute of Science, Bengaluru, India.


\begin{thebibliography}{0}

	\bibitem{B72} J. D. Bekenstein, {\em Nuovo Cimento, Lett.} {\bf 4}, 773
	(1972).
	\bibitem{BK74} G. S. Bisnovatyi-Kogan, A. A. Ruzmaikin, {\em Ap\&SS} {\bf 28}, 45
	(1974).
	\bibitem{TNM11} A. Tchekhovskoy, R. Narayan, J. C. McKinney, {\em MNRAS} {\bf 418}, L79 
	(2011).
	\bibitem{NIA03} R. Narayan, I. V. Igumenshchev, M. A. Abramowicz, {\em PASJ} {\bf 55}, L69
	(2003).
	\bibitem{MKU01} D. L. Meier, S. Koide, Y. Uchida, {\em Science} {\bf 291}, 84
	(2001).
	\bibitem{MM19a} T. Mondal, B. Mukhopadhyay, {\em MNRAS} {\bf 482}, L24
	(2019).
	\bibitem{MM19b} T. Mondal, B. Mukhopadhyay, {\em MNRAS} {\bf 486}, 3465
	(2019).
	\bibitem{Z14} M. Zamaninasab, et al., {\em Nature} {\bf 510}, 126 
	(2014).
	\bibitem{B02} B. Mukhopadhyay, {\em ApJ} {\bf 581}, 427
	(2002).
	\bibitem{SS73} N. Shakura, R. Sunyaev, {\em A\&A} {\bf 24}, 337
	(1973).
	\bibitem{C96} S. K. Chakrabarti, {\em ApJ} {\bf 464}, 664
	(1996).
	\bibitem{MM18} T. Mondal, B. Mukhopadhyay, {\em MNRAS} {\bf 476}, 2396
	(2018).
	\bibitem{DM19} S. R. Datta, B. Mukhopadhyay, {\em MNRAS} {\bf 486}, 1641
	(2019).
	\bibitem{BH98} S. A. Balbus, J. F. Hawley, {\em Rev. Mod. Phys.} {\bf 70}, 1
	(1998).
	\bibitem{MC15} B. Mukhopadhyay, K. Chatterjee, {\em ApJ} {\bf 807}, 43
	(2015).
	\bibitem{MMMG15} T. Mondal, B. Mukhopadhyay, {\em this volume}.
	\bibitem{MMG15} B. Mukhopadhyay, {\em this volume}.

\end{thebibliography}
\end{document}